# Two pathways to break the insulating state in a correlated transition metal oxide

Joel Kuttruff,[1] Ritwika Mandal,[2] Marina Servol,[2, 3] Céline Mariette,[4] Hiroko Tokoro,[3,5] Shin-ichi Ohkoshi,[3,6] Rodolphe Sopracase,[1] Hervé Cailleau,[2, 3] Laurent Cario,[7, 3] Etienne Janod,[7, 3] Maciej Lorenc,[2, 3] and Vinh Ta Phuoc[1, 3]

[1]*GREMAN - UMR7347 CNRS, Université de Tours, INSA Centre Val de Loire, Parc de Grandmont, 37200 Tours - France*
[2]*Université de Rennes, CNRS, IPR (Institut de Physique de Rennes) - UMR 6251, F-35000 Rennes, France*
[3]*DYNACOM IRL2015 University of Tokyo - CNRS - UR1, Department of Chemistry, 7-3-1 Hongo, Tokyo 113-0033, Japan*
[4]*European Synchrotron Radiation Facility, 71 avenue des Martyrs, Grenoble, 38043, France*
[5] *Department of Materials Science, Institute of Pure and Applied Sciences, University of Tsukuba, Tsukuba, Ibaraki, Japan*
[6] *Department of Chemistry, School of Science, The University of Tokyo, Bunkyo-ku, Tokyo, Japan*
[7]*Nantes Université, CNRS, Institut des Matériaux Jean Rouxel, IMN, UMR 6502, F-44000 Nantes, France*

**Correlated transition metal oxides present exciting prospects as switches or memory and storage devices owing to the possibility to control electronic properties using various external stimuli. While their complex behaviour is known to stem from interplay between electronic correlations, atomic structure and orbital physics, they remain poorly understood on the microscopic level. Here, we investigate such origins as a function of temperature and pressure in the transition metal oxide $Ti_3O_5$. We find that the insulating room-temperature phase is characterized by one-dimensional zig-zag chains composed by two types of titanium dimers forming orbital selective valence bonds. At the thermal phase transition, one type of titanium dimer breaks up, resulting in an insulator to metal transition with a large orbital repopulation between the two states. Moreover, optical spectroscopy reveals that an additional pressure-driven insulator to metal transition occurs in $Ti_3O_5$ at room temperature. The phenomenology of this novel pressure-induced metallic transition is completely different from the insofar studied transitions and results from a competition between intra- and inter-dimer hopping. Our combined results suggest that $Ti_3O_5$ is a prototypical correlated transition metal oxide, where both correlations as well as orbital interactions need to be considered to fully understand the evolution of the electronic states.**

The exploration of transition metal oxides with tunable electronic properties has become a key area of research, owing to their potential applications in a wide range of technologies, including energy storage and management [1-3], catalysis [4], or quantum devices [5]. In particular, titanium oxides, with their complex electronic structures and the ability to undergo reversible phase transitions, present an exciting opportunity for engineering materials with tailored electronic and optical responses. Among these, the titanium oxide $Ti_3O_5$ stands out due to its polymorphism which leads to distinct electronic behavior and makes it prone to various external stimuli [6-18]. In this context, $Ti_3O_5$ shows strong potential in applications like solar steam generation [13] or heat storage [8]. $Ti_3O_5$ undergoes a thermal phase transition between a non-magnetic insulator (so-called β) and a paramagnetic metallic (so-called λ) phase around $T_{IM}$=460 K [19]. This insulator-to-metal transition is isosymmetric (i.e., maintaining the monoclinic space group C2/m, Fig.1), markedly first order, and characterized by significant changes in volume (+6.4%) and latent heat [7]. At $T_{st}$=500 K, a second structural phase transition takes place to a high-symmetry metallic α-$Ti_3O_5$ phase (space group Cmcm), with only small changes of transport and magnetic properties [6]. Most of the recent theoretical and experimental effort has focused on $Ti_3O_5$ in nano-crystalized, nano-granular or thin-film forms [20-24]. For those morphologies, the λ-phase is stable down to room temperature and below, with further possibilities to tune the electronic structure and phase transition by doping [25-29] and enabling, for instance, photo-reversible switching between β- and λ-phases [7]. Time-resolved X-ray diffraction experiments have suggested that the light-induced transition is induced by the propagation of an acoustic strain wave associated to a volume expansion [11], although the microscopic origin of the transition is still under debate [30].

The electronic structure of $Ti_3O_5$ nano crystals has been studied by hard X-ray photoelectron spectroscopy [9]. While those experiments suggest a dominant contribution of electronic correlations within $Ti_3O_5$, the electronic properties also depend critically on the morphology of the sample [21,24,31,32]. In order to exclude the finite size effects of the nano crystals and the substrate effects in thin films, experiments on free-standing single crystals are required. In particular, it has to be clarified whether the intrinsic correlation effects and ion pair formation play a role in the formation of the low-temperature insulating phase as well as its breakup into a metallic state beyond the phase transition.

Here, we address this question with broadband optical spectroscopy experiments as a function of temperature and pressure, as well as density functional theory (DFT) calculations. We show that β-$Ti_3O_5$ is a narrow-gap insulator characterized by two kinds of titanium dimers, forming orbital selective valence bonds. The thermal insulator to metal transition, related to the first order isosymmetric structural transition, is due to dissociation of one of the dimers, associated to a large orbital repopulation between the two states, the sudden appearance of a narrow Drude peak, and a large transfer of spectral weight. The expansion along the *c*-axis upon the phase-transition suggests that external pressure can be a control parameter for the electronic properties of $Ti_3O_5$. Following this route, we find a novel pressure-driven insulator-metal transition starting from the β-phase which does not seem to be concomitant to any structural symmetry break. This transition appears to be gradual and drastically differs from the thermal β–λ transition. The resulting high-pressure phase exhibits a metallic behavior only along the $c$ crystallographic axis resulting from a competition between bandwidth increase and bandspliting into bonding and antibonding orbitals.

Figure 1 shows the optical conductivity and structure of single-crystal $Ti_3O_5$ as a function of temperature. Figure 1a illustrates the crystal structure for β-, λ-, and α-phases. The unit cell (see Supplementary Fig. S1) contains 6 Ti atoms in 3 different Wickoff positions denoted $Ti_{1-3}$. All transitions include structural changes of the crystal, which are dominant along the crystal $c$-axis. In addition, while the β-to-λ transition is isosymmetric, the λ-to-α transition also involves a transformation from monoclinic to orthorhombic structure. Figures 1b-c show reflectivity measurements for light polarized along the $a$-axis (Fig. 1b) and $c$-axis (Fig.1c), respectively, as a function of temperature across the phase-transitions introduced above. While the observed anisotropy is relatively low, we see strong changes of reflectivity across the transitions as a function of temperature in the whole measured spectral region. From the reflectivity spectra, we calculate the optical conductivity (Fig. 1d-e) by Variational Dielectric Function (VDF) analysis [33]. We find that β-$Ti_3O_5$ is an insulator at room temperature with an optical gap of ~300 meV for both E∥$a$ and E∥$c$ polarizations, in agreement with a previous report from Saiki et al. [32]. The sharp peaks below 100 meV are due to phonon modes and show a strong anisotropy. The number of observed modes is consistent with Group Theory (see Supplementary Fig. S2). The gap shows only a weak temperature dependence below the transition (Supplementary Fig. S3). However, it collapses at $T_{IM}$ = 460 K and a narrow Drude peak suddenly appears mostly along the $a$-axis, concomitant with a small decrease of the optical conductivity in the mid-infrared visible range and in accordance with transport measurements [6]. In addition, a strong screening of phonon modes due to metallicity is observed in reflectivity spectra above $T_{IM}$.

To further investigate the electronic structure and assign optical excitations, we performed Density Functional Theory calculations (see "Methods"). The results of our calculations show an insulating state in the low-temperature β-phase with a direct gap of ~300 meV at Γ (Supplementary Fig. S4), in agreement with previous calculations [7,34] and our experimental results (see Fig. 1). Around the Fermi level, the electronic structure of the β-phase mainly consists of Ti-3$d$ bands (Supplementary Fig. S4), as also observed in other early 3d transition metal oxides [35-37]. To give a comprehensive picture of the electronic structure, we calculated densities of states projected onto cubic harmonics using a local coordinate system for each titanium atom. Figure 2a shows the result of this projection on $d_{xy}, d_{xz}, d_{yz}, d_{x2-y2}$, and $d_{z2}$ harmonics. Octahedral crystal field splits 3d levels into t2g (from ~-2 eV to ~2 eV) and eg bands (above ~2 eV). In addition, the degeneracy is lifted within the t2g manifold. Indeed, only $d_{xz}$ orbitals of $Ti_1$ and $Ti_3$ atoms are occupied forming nearly flat bands at -1.3 eV and -0.6 eV, respectively. In contrast, the $Ti_1$ and $Ti_3$ $d_{xy}$ and $d_{yz}$ orbitals are above the Fermi level, and all $Ti_2$-d orbitals remain almost empty (3d0). The bands just above the Fermi level are mostly of $Ti_2$ character (Supplementary Fig. S4). Furthermore, the $Ti_1$ and $Ti_3$ $d_{xz}$ orbitals split into bonding and antibonding molecular orbitals, resulting in the formation of $Ti_1$-$Ti_1$ and $Ti_3$-$Ti_3$ dimers associated with short metal-metal distances (σ-type bonding). This splitting is larger than 3 eV for both dimers. Figure 2b shows the corresponding charge density for bands around the Fermi level. The dimerisation can clearly be seen in the charge density, forming 1D zig-zag chains along the crystallographic c-axis. In these chains, although their respective weights strongly differ, both $Ti_1$ and $Ti_3$ $d_{xz}$ orbitals contribute to occupied bands at -1.3 eV and -0.6 eV, indicating non-zero hopping-integrals between adjacent molecular orbitals. Note that the formation of such orbital-selective valence bonds is fully compatible with the non-magnetic insulating character of β-$Ti_3O_5$ experimentally observed [6]. An atomic picture of this valence bond formation and the resulting band structure scheme are summarized in Fig. 2c. There are six Ti atoms in the unit cell, with a formal valence 3d0.67. Only the $d_{xz}$ orbitals are pointing towards each other along the short $Ti_1$-$Ti_1$ and $Ti_3$-$Ti_3$ distances. This overlap results in the formation of a σ-type chemical bonding between the $Ti_1$-$Ti_1$ and $Ti_3$-$Ti_3$ atoms. Consequently, the t2g manifold splits into bonding (B) ($Ti_1$-σ($d_{xz}$) and Ti3-σ($d_{xz}$)) and antibonding (AB) (Ti1-σ*($d_{xz}$) and Ti3-σ*($d_{xz}$)) molecular orbitals, and non-bonding (NB) ($d_{xy}$ and $d_{yz}$) atomic orbitals. The two bonding molecular orbitals are occupied by the four available d electrons. All other t2g orbitals are empty. Such picture indicates a more structural component of the insulating gap stemming from dimer formation and points away from a purely correlation-driven Mott-Hubbard scenario that was suggested for instance by Fu et al. [38]. Instead of directly driving the insulating state via band splitting, here the role of electronic correlations is to enforce large enough localization of electrons within dimers that in turn prevents conduction [34]. In that sense, β-$Ti_3O_5$ can be regarded as a correlated, but structurally driven insulator. We note that the physics of the β-phase of $Ti_3O_5$ recalls the widely studied and highly debated transition metal oxide $VO_2$ [39-46] where the insulating phase is related to the formation of V dimers along chains in the c-direction of the rutile structure.

To compare with experiments, we calculate the optical conductivity tensor from our simulation results (Fig 3). For this calculation, we use cartesian axes where x, y, and z are roughly aligned along the crystallographic a, c, and b axis, respectively. Figures 3a-d show optical conductivity spectra of β-$Ti_3O_5$ measured at 300 K (black curves in Fig. 3a,c) compared to simulation results (red dots in Fig. 3b,d) and indicate good agreement overall between theory and experiment. To fit the experimental data, we use three Lorentz oscillators for both polarizations. The two mid-infrared oscillators are centered around 0.9 eV (blue curve) and 1.6 eV (green curve), and an additional oscillator (yellow curve) mimics high energy features. Similarly, the calculated optical conductivity tensor for the β-phase shows that the mid-infrared spectrum mainly consists of two transitions: from the occupied $Ti_1$- $Ti_1$ band (at -1.3eV) and from the occupied $Ti_3$-$Ti_3$ band (at -0.6 eV) to unoccupied states (mostly Ti2 $d_{xz}$ orbitals) [11]. Note that $\sigma_{zz}$ mainly consists of excitations from $Ti_3$- $Ti_3$ dimers while contributions arising from both $Ti_1$-$Ti_1$ and $Ti_3$-$Ti_3$ dimers are observed in $\sigma_{xx}$, in agreement with the fit of the experimental data. The $\sigma_{yy}$ tensor element is almost zero up to 3 eV (see Supplementary Fig. S5), in agreement with a previous report by Saiki et al. [32].

Figures 3e-f show corresponding optical conductivity spectra above the insulator-metal transition for λ-$Ti_3O_5$. Experimentally, we observe a decrease of mid-infrared spectral weight and appearance of a narrow band around 0.6 eV (purple) for both polarizations. In addition, a narrow Drude component is essential to fit experimental spectra at least along the *a*-axis ($\sigma_{xx}$, Fig. 3e), evidencing the transition from an insulating to a metallic state. Also for the λ-phase, we find a reasonable agreement between experiments and DFT calculations. Projected DOS for λ-$Ti_3O_5$ are shown in Figure 4 (see Supplementary Fig. S6 for the full band structure and Supplementary Fig. S7 for charge densities). Both Drude peak and the 0.6-eV band arise from a large charge redistribution from $Ti_3$ $d_{xz}$ to $Ti_2$ *d*-orbitals, as seen from projected DOS in Fig.4a. Indeed, the expansion along the *c*-axis is associated with a large rotation of $Ti_3$-$Ti_3$ pairs (see Supplementary Fig. S8). The local environment of both $Ti_2$ and $Ti_3$ drastically changes and the $Ti_3$-$Ti_3$ dimers dissociate. Conversely, the $Ti_1$-$Ti_1$ dimers remain almost unaffected. The bands at the Fermi level are due to $Ti_2$ and $Ti_3$ $d_{xy}$ and $d_{yz}$ orbitals as evidenced by the charge density (Fig.4b). An atomistic description is depicted in Fig. 4c. More in detail (see also Supplementary Fig. S6), the dispersion along C2-Y2 direction of the band at the Fermi level is drastically increased compared to the β-phase. As a consequence, the metallicity mostly appears along the *b*-axis and to a lesser extent along the *a*-axis (see the conductivity tensor elements in Supplementary Fig. S5b) which explains that only a small Drude contribution is experimentally observed along the *a*-axis ($\sigma_{xx}$). Thus, $Ti_3O_5$ undergoes a transition from a valence bond insulator, forming zig-zag chains along *c*-axis, to a partially dimerized metallic state within the *ab* plane, concomitant with a large charge transfer. Similarly to $VO_2$ [44-46], both Mott and Peierls physics are at play in the insulating state as well as the high temperature metallic phase of $Ti_3O_5$.

Since both orbital and electronic correlations are known to be highly sensitive to interatomic distances [47], applying an external pressure provides a powerful method to continuously tune these effects and probe the limits of the β-$Ti_3O_5$ electronic state. According to Raman spectroscopy and X-ray diffraction measured up to tens of GPa, no structural phase transition was detected up to ~38 GPa [48,49]. We measure the unpolarized reflectivity as function of pressure at room temperature using the diamond anvil cell technique. Figure 5a shows the results of this measurement and Figure 5b the transformation to optical conductivity with VDF analysis. A large increase of low energy reflectivity is observed as pressure is increased, indicating a clear metallization. Concomitantly, the only phonon mode observed in our spectral range around 100 meV exhibits a clear blue-shift due to unit cell shrinking. Optical conductivity also exhibits a large spectral weight transfer from the mid-infrared to lower energies at high pressure. Such a transfer is accompanied with the appearance of a Drude peak and a narrow band around 0.3 eV which progressively merges with the Drude peak (Fig. 5b). We observe a closing of the optical gap around ~4-5 GPa, related to a pressure-induced insulator-metal transition in $\beta$-$Ti_3O_5$. Note that the low-energy conductivity of this pressurized $\beta$-$Ti_3O_5$ phase at 16 GPa is about three times larger than in the $\lambda$-phase. Moreover, the evolution of optical quantities across the transition is continuous in contrast with the thermal transition. Figure 5c shows the evolution of spectral weight at two selected energies in the visible (2.2 eV, blue squares) and mid-infrared (0.5 eV, red quares) region, respectively. Interestingly, the low-energy spectral weight gradually increases as function of pressure while the high energy spectral weight rapidly decreases and remains mostly constant above the transition (Fig. 5c), indicating a complex evolution of the electronic structure, with mid-infrared spectral weight probably transferred to both low and high energy.

Such a pressure-induced insulator-metal transition without any structural phase transition is usually encountered in strongly correlated systems such as Mott insulators or antiferromagnetic insulators [50-53]. On the contrary, in some systems, increasing the pressure can also promote the formation of di/trimers due to shortening of direct metal-metal distances, resulting ultimately in a pressure-induced metal to insulator transition such as in heavy fermion system $LiV_2O_4$ [54] or Kitaev materials [55,56]. To investigate the role of pressure on the electronic properties of $\beta$-$Ti_3O_5$, we calculate the electronic structure and the optical conductivity at 5, 10 and 15 GPa using the cell parameters experimentally obtained by XRD under pressure [49]. Figure 5d shows the calculated optical conductivity as function of pressure. The metallization and the low-energy electrodynamics are qualitatively well described at the DFT+U level. Figure 6a shows details of the band structure around $\Gamma$ at various pressures (see Supplementary Figs. S9-S11 for full bandstructure and projected densities of states). With increasing pressure, the direct gap at $\Gamma$ between the top of the valence band (of $Ti_3$ character) and the bottom of the valence band (of $Ti_2$ character) decreases due to bandwidth inrease. At 5 GPa, the gap is 0.1 eV and is fully closed at 10 GPa. The calculation at 15 GPa shows an overlap of bands with $Ti_1$ and $Ti_3$ character at the Fermi energy, indicating that $\beta$-$Ti_3O_5$ undergoes a semiconductor to metal transition at an intermediate pressure. Such an overlap transition explains the increase of the low-energy optical conductivity with increasing pressure. Consequently, the $\sigma_{zz}$ (along the zig-zag dimer chains) term of the calculated optical conductivity tensor shows a metallic behavior (see Fig. 5b) while $\sigma_{xx}$ and $\sigma_{yy}$ still display an insulating behavior. Note also that the calculated conductivity for the pressurized β-phase within the *ac* plane is larger than the one of the λ-phase, in agreement with the experiments.

While the temperature and pressure-induced insulator-to-metal transitions are both isosymmetric and associated to a large spectral weight transfer to both low and high energies, their microscopic origins drastically differ. Indeed, the metallicity mostly appears along *b*-axis for the thermal transition, associated to the dissociation of one of the two dimers of the β-phase and a large orbital repopulation. In contrast, this new pressure-driven metallic β-phase results from the competition between bonding- antibonding splitting and the bandwidth enlargement without major modifications of the crystal structure. The former thermal insulator-to-metal transition is controlled by intra-dimer hopping, while the latter pressure-driven one is controlled by inter-dimer hopping. This mechanism and the corresponding schematic band structure are represented in Fig. 6c-d. Note that the decrease of the near-infrared conductivity and the associated spectral weight transfer to high energy observed experimentally is not well reproduced at the DFT+U level. Such a feature is a usual hallmark of strong electronic correlations. It probably would require a more advanced theoretical framework such as DFT and cluster-Dynamical Mean Field Theory to take into account both strong electronic correlations and dimerization [44]. Again, this metallic β-$Ti_3O_5$ phase recalls the physics of $VO_2$, where a transition from a monoclinic insulator to a monoclinic metallic phase at high pressure is also observed [57-61]. However, while for $VO_2$ the critical pressure to induce metallicity is ~10 GPa [57], our results indicate a transition already at ~4 GPa.

In conclusion, we used a combination of optical spectroscopy as function of temperature and pressure with first-principle calculations to study the electronic properties of $Ti_3O_5$ across its various phase transitions. The thermal insulator to metal transition is concomitant with the first-order structural transition and is characterized by the appearance of a small Drude peak mostly along one axis and a large spectral weight transfer from mid-infrared to both low and high frequency in optical conductivity spectra, indicating strong modifications of the electronic structure. DFT+U calculations correctly predict that the room temperature β-$Ti_3O_5$ phase is an insulator where electrons are localized on two types of dimers formed by titanium pairs. The metallicity of the λ-$Ti_3O_5$ is due to the dissociation of one type of dimer resulting in a large orbital repopulation. The calculated optical conductivity spectra show an overall good agreement with experiments for both phases and allow to assign optical transitions. By applying high pressure, we found a new metallic β'-$Ti_3O_5$ monoclinic phase with a larger optical conductivity at low energy than both high-temperature phases. Combining with first-principles calculations, we ascribe this transition to a competition between dimer formation and bandwidth enlargement driven by pressure. Our results highlight the crucial role of orbital effects on electronic properties of $Ti_3O_5$ and demonstrate that external pressure can be used as a practical tool to control its optical properties. A striking feature is that both transitions are isosymmetric although their microscopic mechanisms are completely different. In that sense, our work suggests $Ti_3O_5$ as a formidable playground to study the complex interplay between orbital physics, symmetry and electronic correlations. Understanding this interplay will provide opportunities to tune the electronic properties of transition metal oxides towards various applications, including energy management, ultrafast electronics, or catalysis.

**Acknowledgments:** JK acknowledges support by the Deutsche Forschungsgemeinschaft via SFB 1432. Agence Nationale de la Recherche is acknowledged for financial support under grant number ANR-23-CE30-0027 ('FASTRAIN'). HT and SO acknowledges support by the JST FOREST Program (JPMJFR213Q) and the JST Advanced Technologies for Carbon-Neutral (JPMJAN23A2)

**Data availability statement:** All data is available from the corresponding author upon reasonable request.

**Author contributions:** All authors performed research and wrote the paper.

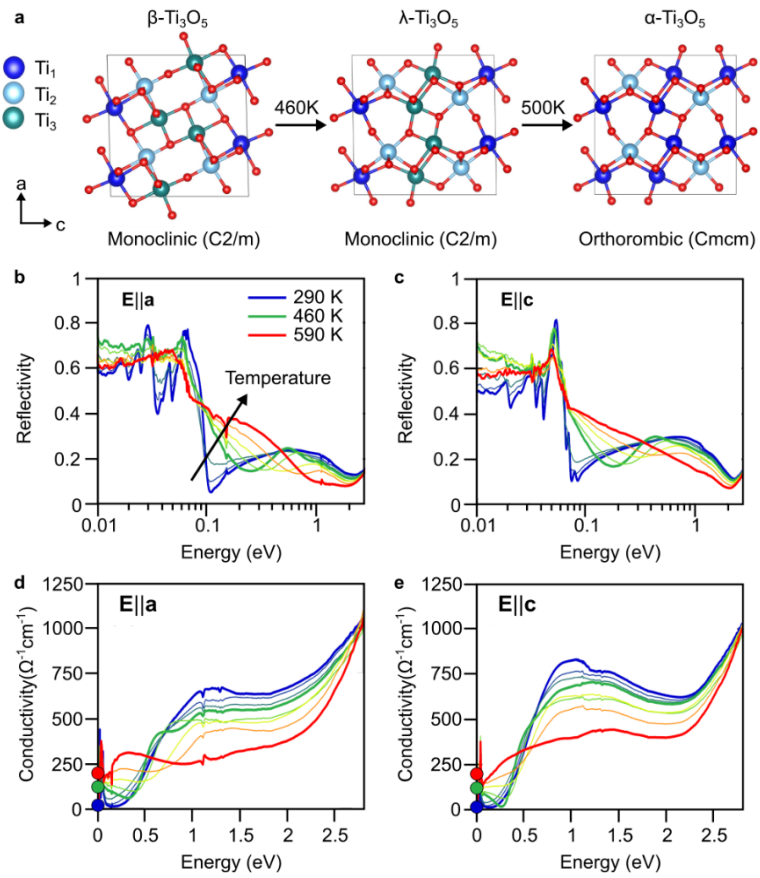

**Fig. 1.** (**a**) Crystal structure of $Ti_3O_5$ as function of temperature. (**b-c**) Reflectivity spectra of $Ti_3O_5$ as function of temperature for polarization along crystallographic *a*-axis (**b**) and *c*-axis (**c**). (**d-e**) Calculated optical conductivity from variational dielectric function analysis of reflectivity measurements. Closed circles are DC conductivity values from transport measurements taken from ref. [6].

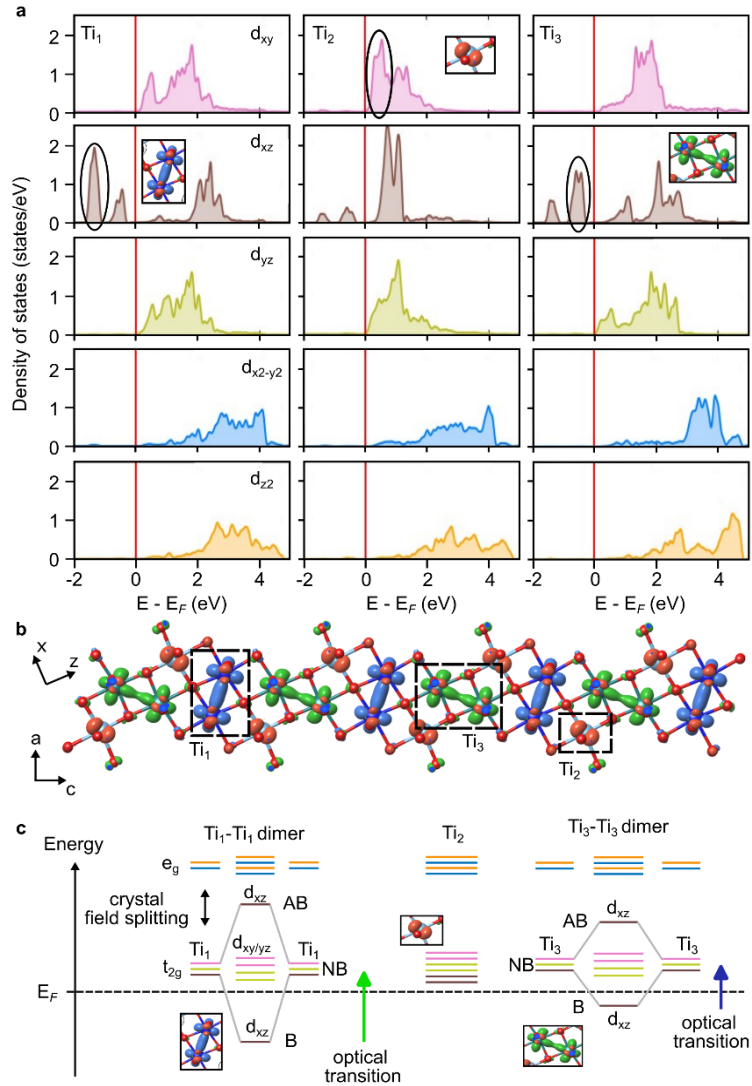

**Fig. 2.** (a) Projected density of states of β-$Ti_3O_5$ on $Ti_1$, $Ti_2$ and $Ti_3$ 3$d$ orbitals showing bonding and antibonding molecular orbitals for both $Ti_1$ and $Ti_3$ arising from $d_{xz}$ atomic orbitals. (b) Charge density from bands around the Fermi level. Crystal axis and cartesian axis used for the orbital-projected density of states are shown on the bottom left and top left, respectively. Charge density indicates that the band at -1.3 eV and -0.6 eV mostly arise from $Ti_1$-$Ti_1$ (blue) and $Ti_3$-$Ti_3$ (green) dimers, respectively, and forms 1D zig-zag chains along the $c$-axis. (c) Illustration of level splitting and formation of nonmagnetic dimers from $t2g$ orbitals. There are six Ti atoms in the unit cell, with a formal valence $3d^{0.67}$. Only the $d_{xz}$ orbitals are pointing towards each other along the short $Ti_1$-$Ti_1$ and $Ti_3$-$Ti_3$ distances. This overlap results in the formation of a σ-type chemical bonding between the $Ti_1$-$Ti_1$ and $Ti_3$-$Ti_3$ atoms. Consequently, the $t2g$ manifold splits into bonding (B) ($Ti_1$-σ($d_{xz}$) and $Ti_3$-σ($d_{xz}$)), antibonding (AB) ($Ti_1$-σ*($d_{xz}$) and $Ti_3$-σ*($d_{xz}$)) molecular orbitals and non-bonding (NB) ($d_{xy}$ and $d_{yz}$) atomic orbitals. The two bonding molecular orbitals are occupied by the four available $d$ electrons. All other $t2g$ orbitals are empty. Expected low-energy optical transitions from the occupied orbitals are denoted by the blue and green arrow, respectively.

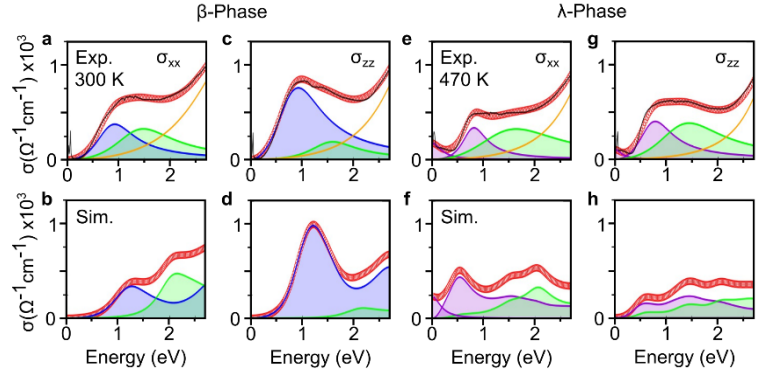

**Fig. 3.** (**a-d**) Comparison between experimental (**a** and **c**) and calculated (**b** and **d**) optical conductivity tensor elements for β-$Ti_3O_5$. Black lines are experimental data and red circles are the sum of fitted Lorentz oscillators (**a** and **c**) and sum of calculated contributions from DFT (**b** and **d**), respectively. The green (blue) curve corresponds to the contribution to optical conductivity arising from $Ti_1$-$Ti_1$ ($Ti_3$-$Ti_3$) dimer band. The orange line corresponds to higher energy contributions. (**e-h**) Same for λ-$Ti_3O_5$. The violet curves, including a narrow Drude peak and a broad mid-infrared peak, are due to bands crossing the Fermi level with a strong $Ti_2$ and $Ti_3$ character.

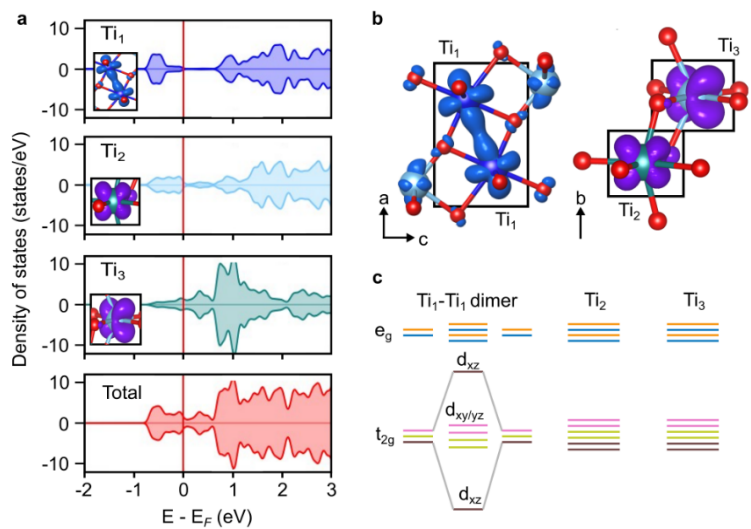

**Fig. 4.** (**a**) Density of states for λ-$Ti_3O_5$ and projected densities of states for *d*-orbitals of $Ti_1$, $Ti_2$ and $Ti_3$ atoms, respectively. (**b**) Charge densities for bands around the Fermi level. (**c**) Illustration of level splitting and formation of nonmagnetic dimers from *t2g* orbitals of $Ti_1$ atoms.

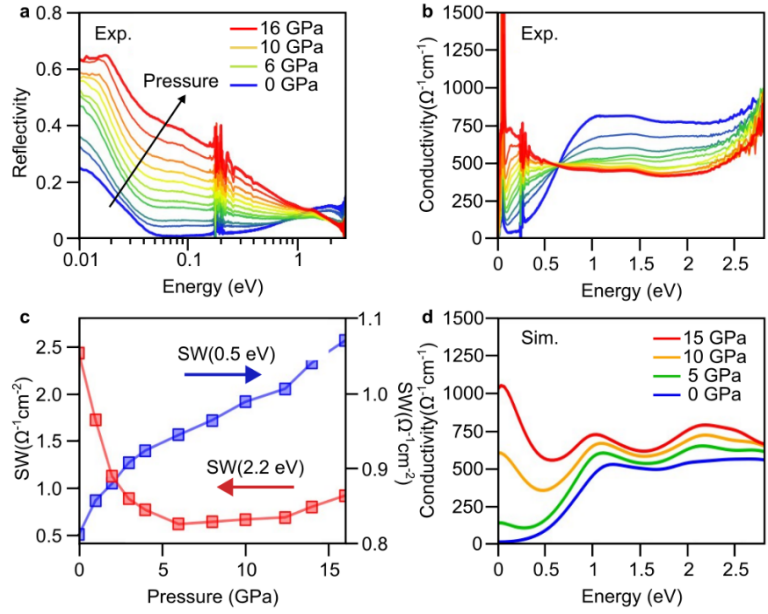

**Fig. 5.** (**a**) Reflectivity in the diamond anvil cell at various pressures. (**b**) Resulting optical conductivity. (**c**) Spectral weight from (**b**) as a function of applied pressure evaluated at 0.5 eV (blue) and (2.2 eV), respectively. (**d**) Calculated optical conductivity spectra of β-$Ti_3O_5$ as a function of pressure obtained from the diagonal terms of the conductivity tensor : $\sigma = (\sigma_{xx}+\sigma_{yy}+\sigma_{zz})/3$.

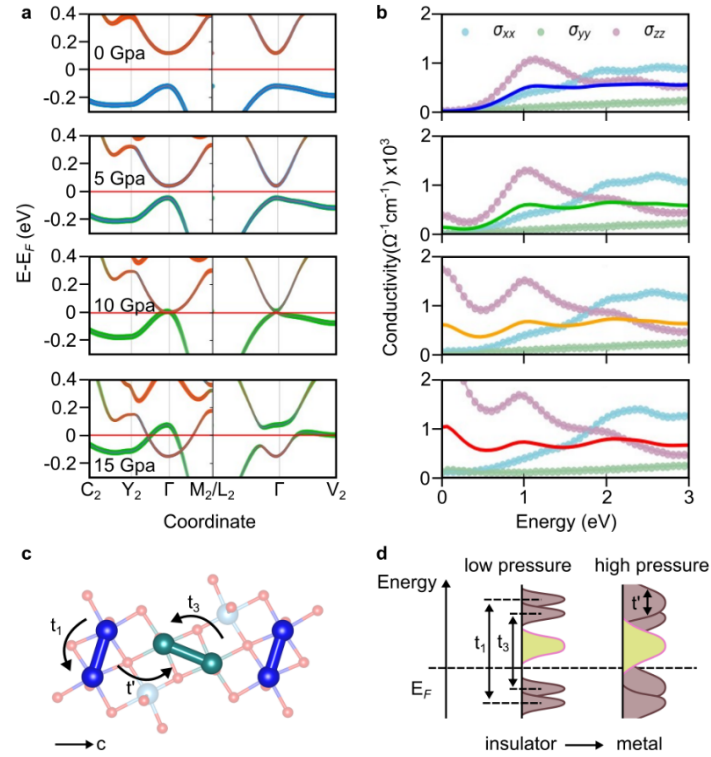

**Fig. 6.** (**a**) Band structure around the Gamma point at 0, 5, 10 and 15 GPa using 'fat-band' representation. The weight of the $3d$ orbital character is shown as the line thickness. The green, orange and blue colors represent $Ti_1$, $Ti_2$ and $Ti_3$ character, respectively. As pressure increases, the color change from blue to green of the $Ti_3$ band below the Fermi Energy $E_F$ indicates an increase of the mixing between $Ti_3$ and $Ti_1$ character. (**b**) Calculated optical conductivity at 0, 5, 10 and 15 GPa. The circles are the diagonal elements of the conductivity tensor, i.e. $\sigma_{xx}$, $\sigma_{yy}$ and $\sigma_{zz}$, and the solid line is their average. (**c**) Representation of intra- and inter- dimer hopping. (**d**) Schematic band structure showing bonding-antibonding splitting, non-bonding orbitals and bandwidth for low (insulator) and high (metal) pressure phases.

## Materials and Methods

**Optical spectroscopy on single crystals**

Temperature dependent polarized optical conductivity spectra have been deduced from nearly normal incidence reflectivity measured between 100 and 25000 cm$^{-1}$ on a 200x200 µm$^2$ single crystal polished up to optical grade. Ti$_3$O$_5$ single-crystal were provided by NICHIA Corporation.. Great care was taken to measure an area without any crack at the insulator-to-metal transition. Indeed, we first measured the reflectivity on several areas of the sample as function of temperature. At the transition, many cracks and deformations appeared on the sample surface. We then chose large enough areas without any crack and reproduced measurements to ensure that the measurements were reproducible with (i) those first obtained at room temperature, (ii) those obtained for several thermal cycles. Room-temperature pressure-dependent optical conductivity spectra have been deduced from nearly normal incidence sample-diamond reflectivity between 550 and 25 000 cm$^{-1}$ on a 100x100 µm$^2$ single crystal from 0 to 17.5 GPa using a BETSA membrane diamond anvil cell. The sample was loaded inside a gasket hole together with KBr as a hydrostatic medium. The gasket was used as a reference mirror. Pressure was measured with the standard ruby fluorescence technique. Spectra were measured by using a homemade high-vacuum microscope including an X15 Schwarzchild objective connected to a BRUKER IFS 66v/S Fourier Transform Spectrometer with a Mercury-Cadmium-Telluride detector, glowbar, and tungsten light sources. In order to obtain the optical conductivity from the reflectivity, we used a variational dielectric function method [33].

**Density Functional Theory**

All calculations have been carried out by using the Quantum ESPRESSO package, with the Perdew-Burke-Ernzerhoff generalized gradient approximation to describe the exchange-correlation functional. In order to account for the on-site Coulomb repulsion among the Ti 3$d$ electrons, a Hubbard parameter $U$ was added to the Local Spin-Density Approximation (LSDA) functional. We chose $U$=3 eV. However, choosing the value of $U$ over a relatively wide range does not qualitatively change our results. A study of $U$ dependence can be found in ref. [34]. Both Projector Augmented Wave basis and norm-conserving pseudo-potentials were used. The Monkhorst-Pack grid of 8 × 8 × 4 in the reciprocal space was used for the Brillouin zone sampling for both λ-phase and β-phase. The total energy of the system converged to less than $1.0 \times 10^{-6}$ Ry. Electronic wave functions were represented in a plane wave basis up to an energy cutoff of at 90 Ry. Crystallographic structures were taken from ref [19] and ref. [49] for ambient and applied pressure, respectively. Antiferromagnetic (AFM) order was considered. For calculations at high pressure, the experimental cell parameters were considered together with a relaxation of the internal atomic coordinates. The optical properties were computed using

epsilon.x post-processing tool of the Quantum Espresso package, at the independent-particle approximation level. Both intraband and interband contributions were considered.

# Supplementary Material

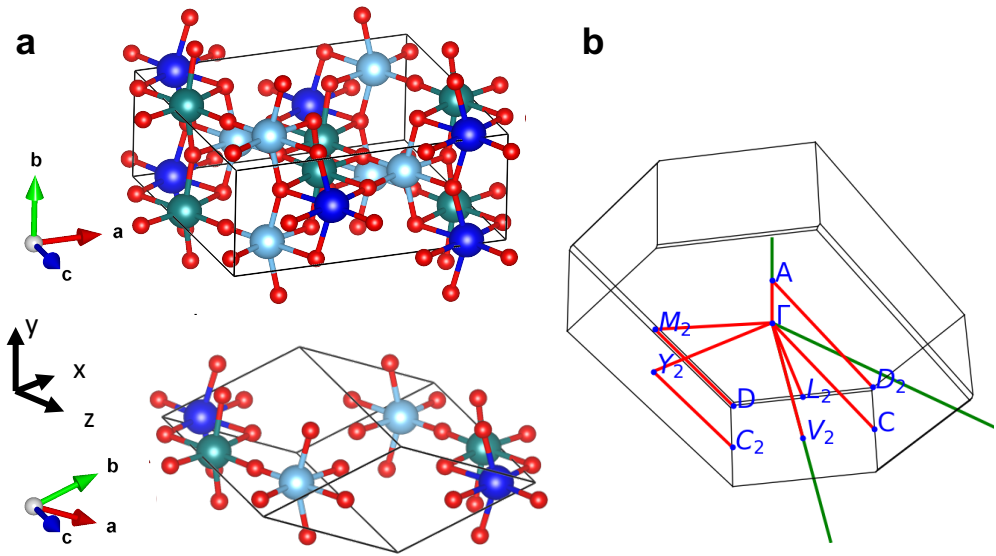

**Fig. S1.** (**a**) Orientation of conventional and primitive unit cells of $\beta$-$Ti_3O_5$ with cartesian axes used for DFT calculations. For projected densitiy of states (PDOS) of Ti atoms, cartesian axes are chosen along octahedral directions. (**b**) Brillouin zone with high-symmetry points used in band structure representation.

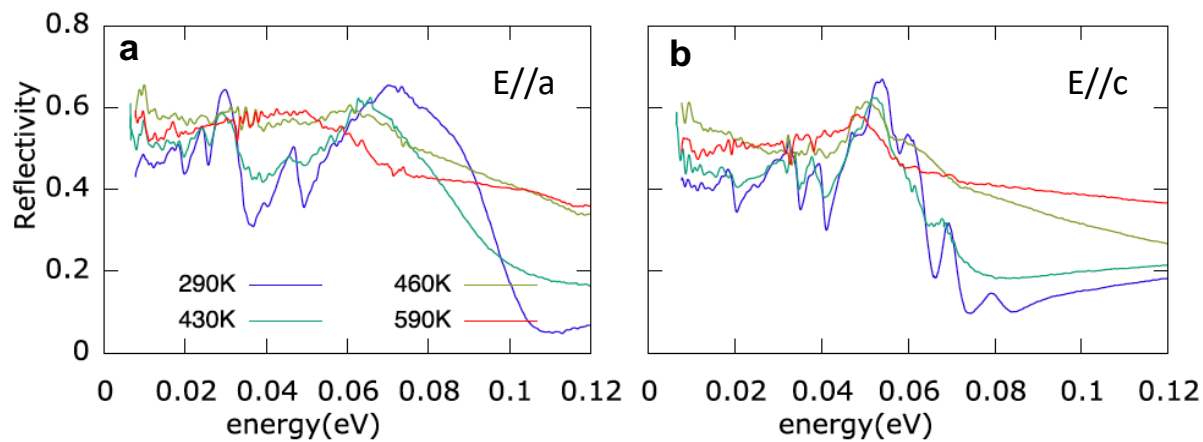

**Fig. S2.** (**a-b**) Reflectivity spectra in the far-infrared spectral range as function of temperature for light polarization along crystal *a*- (**a**) and *c*- (**b**) axis, respectively . 13 of the 14 IR-active phonon modes predicted by Group Theory in *ac*-plane are observed in β-phase. Group theory analysis for C2/m space group (β and λ-phase) predicts 21 IR-active modes / 14 IR-active modes within ac-plane : $\Gamma_{acoustic}$ = $A_u$ + $2B_u$; $\Gamma_{optic}$ = $16A_g$ + $7A_u$ + $8B_g$ + $14B_u$ with $A_u$ and $B_g$ modes along b-axis, $A_g$ and $B_u$ modes within *ac*-plane.

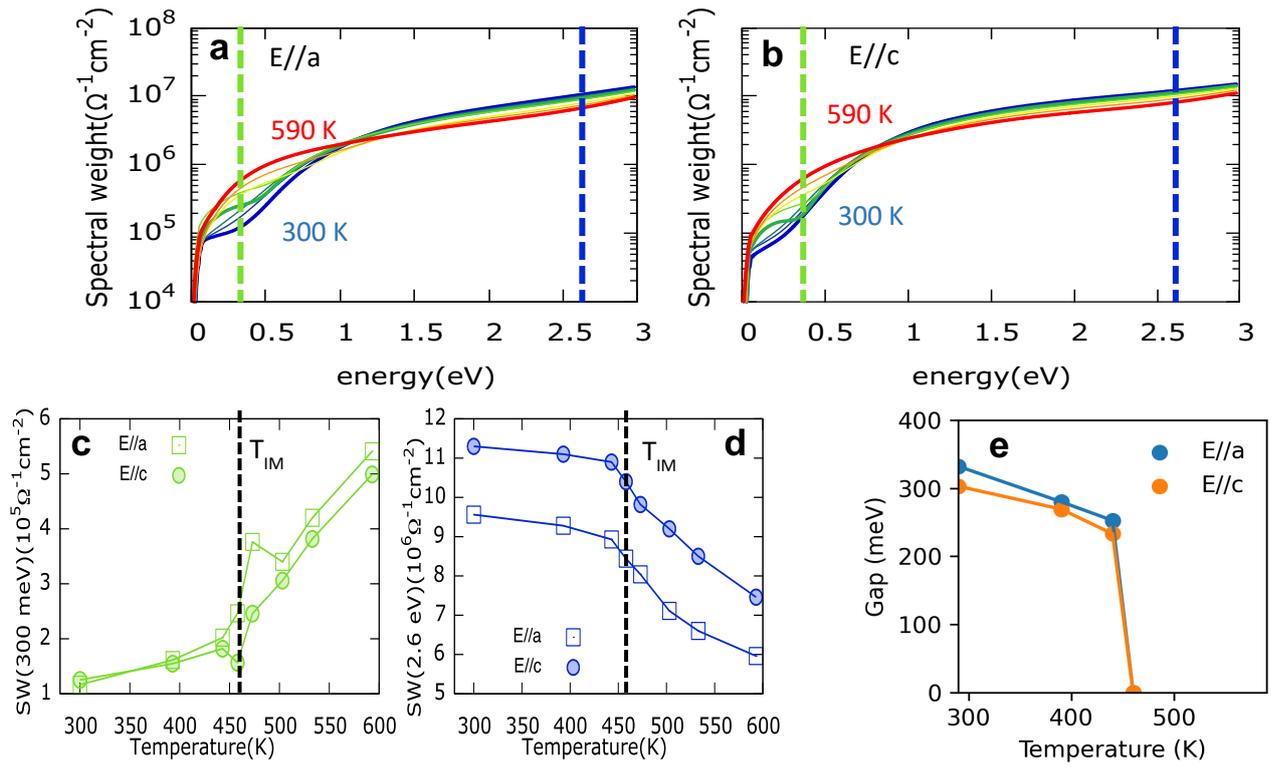

**Fig. S3.** (**a-b**) Spectral weight as function of energy at various temperatures for light polarized along *a*-axis (**a**) and *c*-axis (**b**). (**c-d**) Spectral weight at 300 meV (**c**) and 2.6 eV (**d**). (**e**) Optical gap as function of temperature for both polarizations.

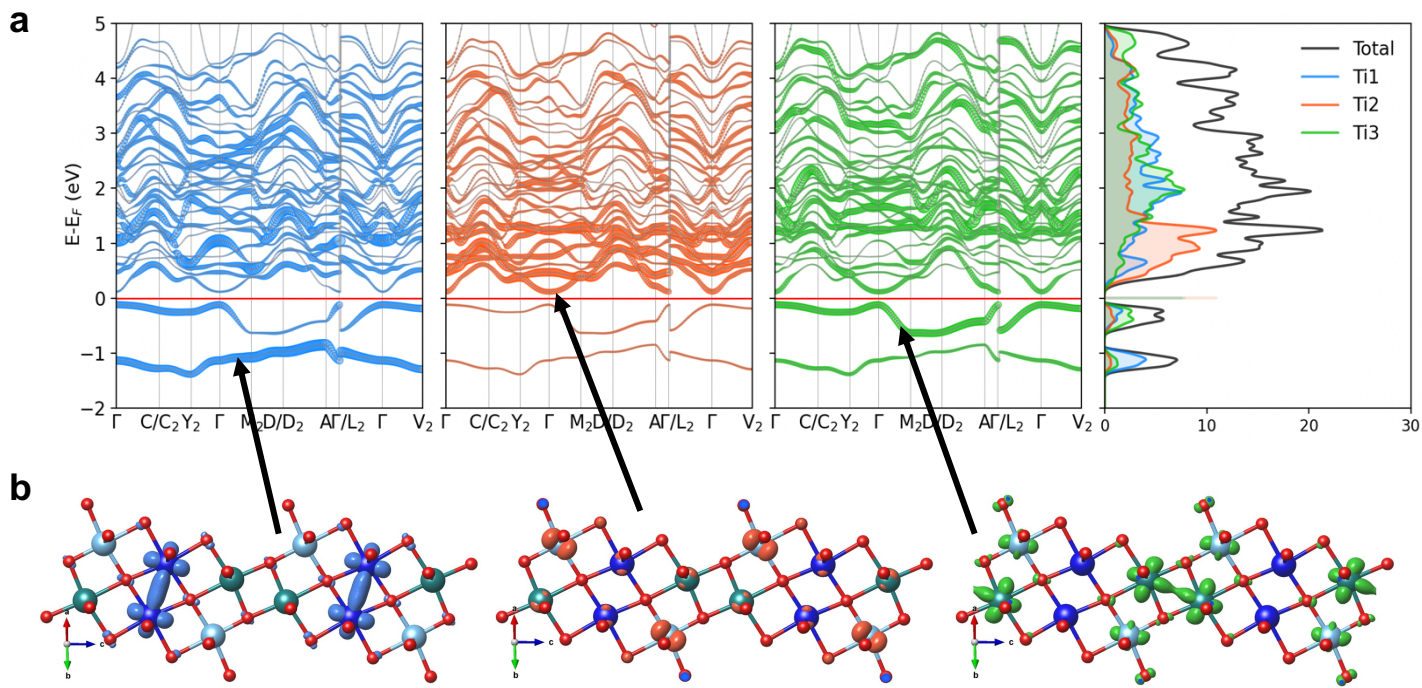

**Fig. S4.** (**a**) Band structure and densities of states of $\beta$-$Ti_3O_5$. Fatbands represent the character of $Ti_1$ (blue - left), $Ti_2$ (orange - center) and $Ti_3$ (green - right) 3d orbitals for each band. The bands at -1eV and -0.2 eV are mostly of $Ti_1$ and $Ti_3$ character, respectively, while the bands just above $E_F$ are mainly of $Ti_2$ character. (**b**) Charge densities for the band just below and above the Fermi level, as indicated by arrows.

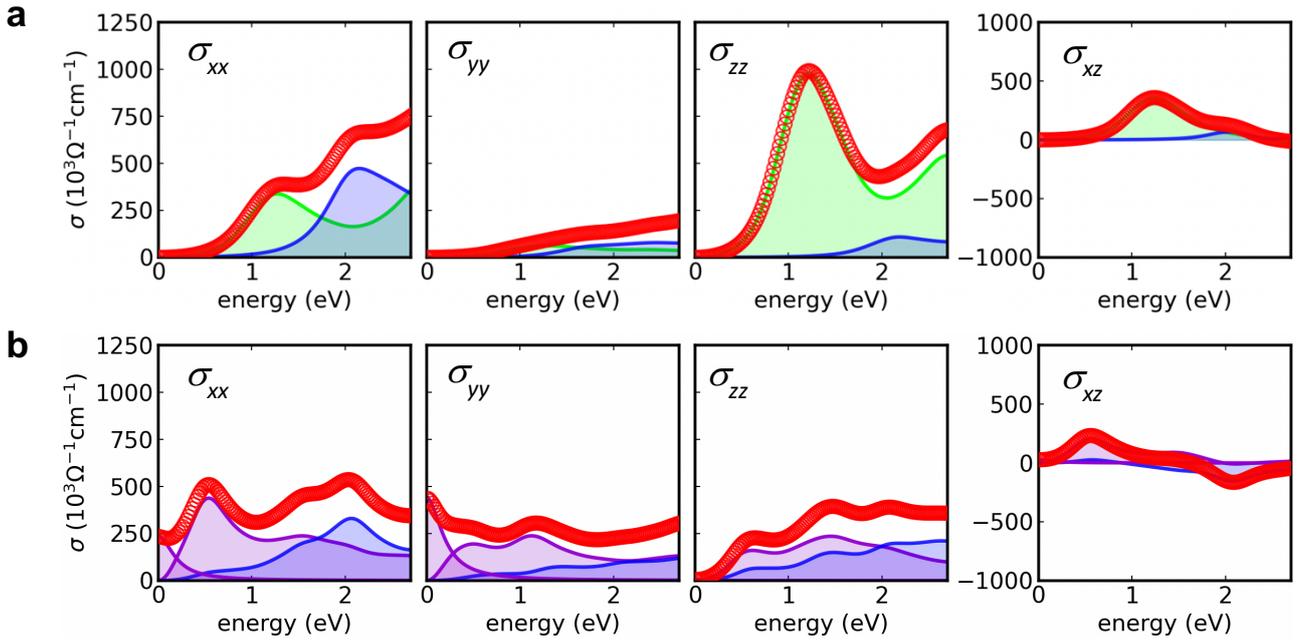

**Fig. S5.** (**a**) Diagonal and off-diagonal elements of the optical conductivity tensor calculated by DFT in $\beta$-Ti$_3$O$_5$ phase. Blue and green contributions arise from Ti$_1$-Ti$_1$ and Ti$_3$-Ti$_3$ dimer bands, respectively. (**b**) Diagonal and off-diagonal elements of the optical conductivity tensor calculated by DFT in $\lambda$-Ti$_3$O$_5$ phase. Blue and purple contributions arise from Ti$_1$-Ti$_1$ dimer and Ti$_2$ + Ti$_3$ bands, respectively.

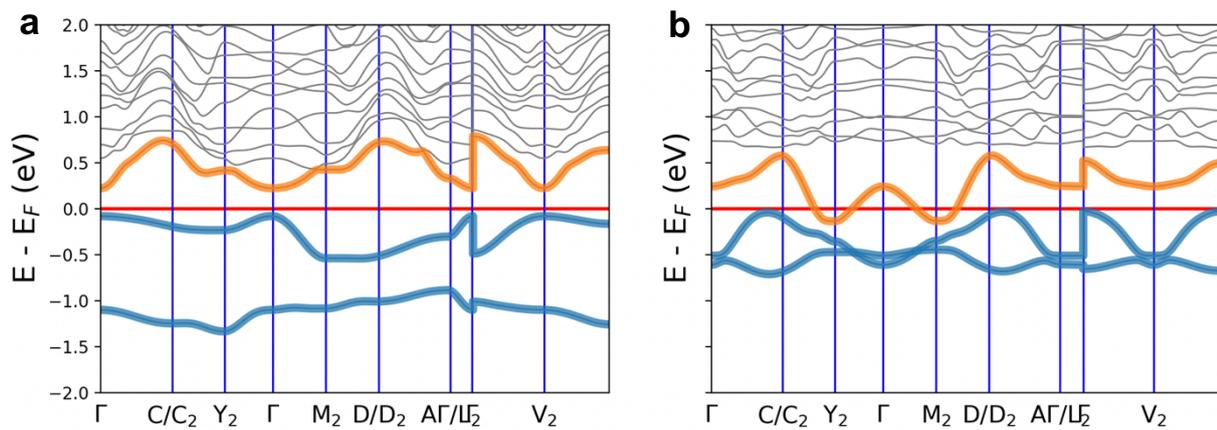

**Fig. S6.** (**a-b**) Comparison of band structures of $\beta$-$Ti_3O_5$ (**a**) and of $\lambda$-$Ti_3O_5$ (**b**). The bands just below (blue) and just above (orange) the Fermi level are highlighted.

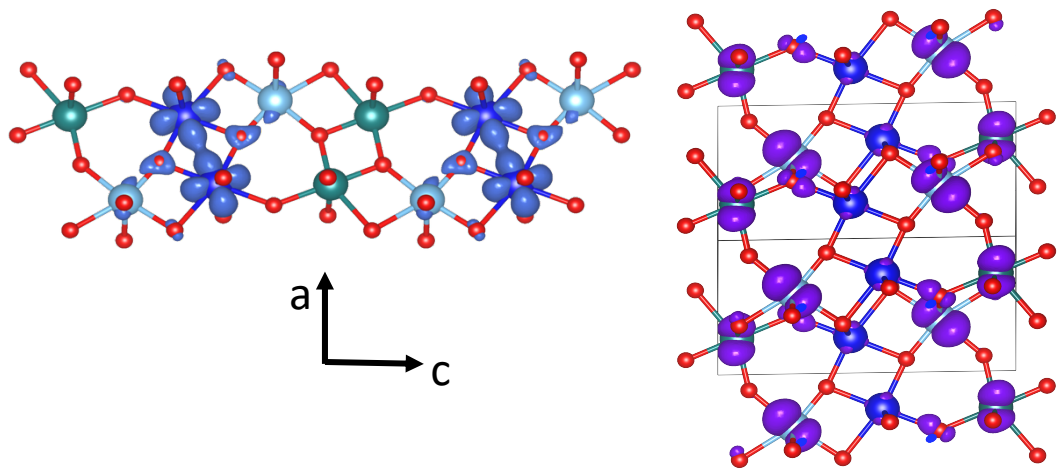

**Fig. S7.** Charge densities for the Ti$_1$-Ti$_1$ dimer band (left), and Ti$_2$ + Ti$_3$ bands (right) around the Fermi level in the $\lambda$-Ti$_3$O$_5$ phase.

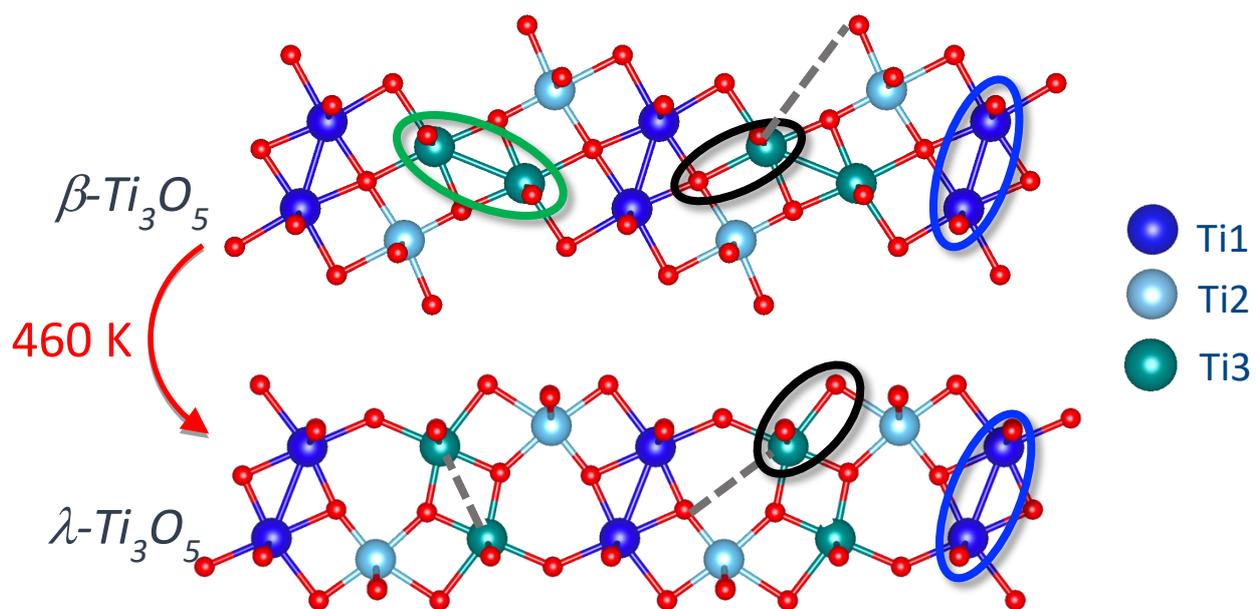

**Fig. S8** Illustration of the structural modifications at the β-Ti₃O₅ to λ-Ti₃O₅ transition at 460K.

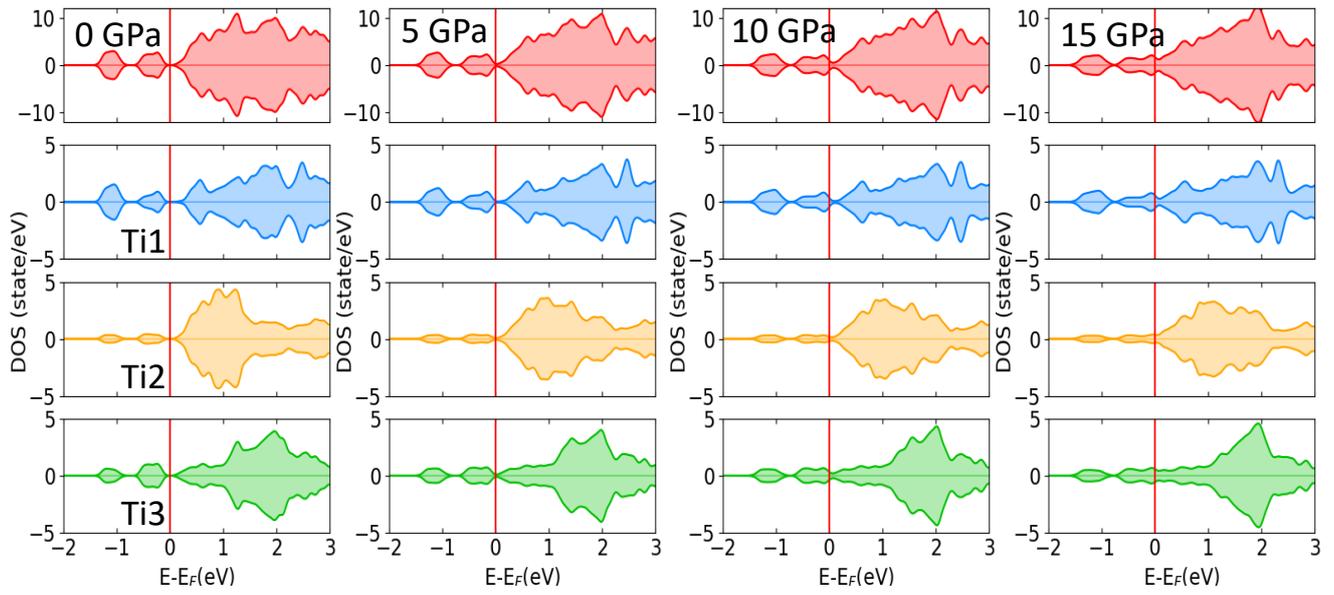

**Fig. S9.** Total (top row) and atom-projected (lower rows) density of states calculated in $\beta$-Ti$_3$O$_5$ phase as function of pressure.

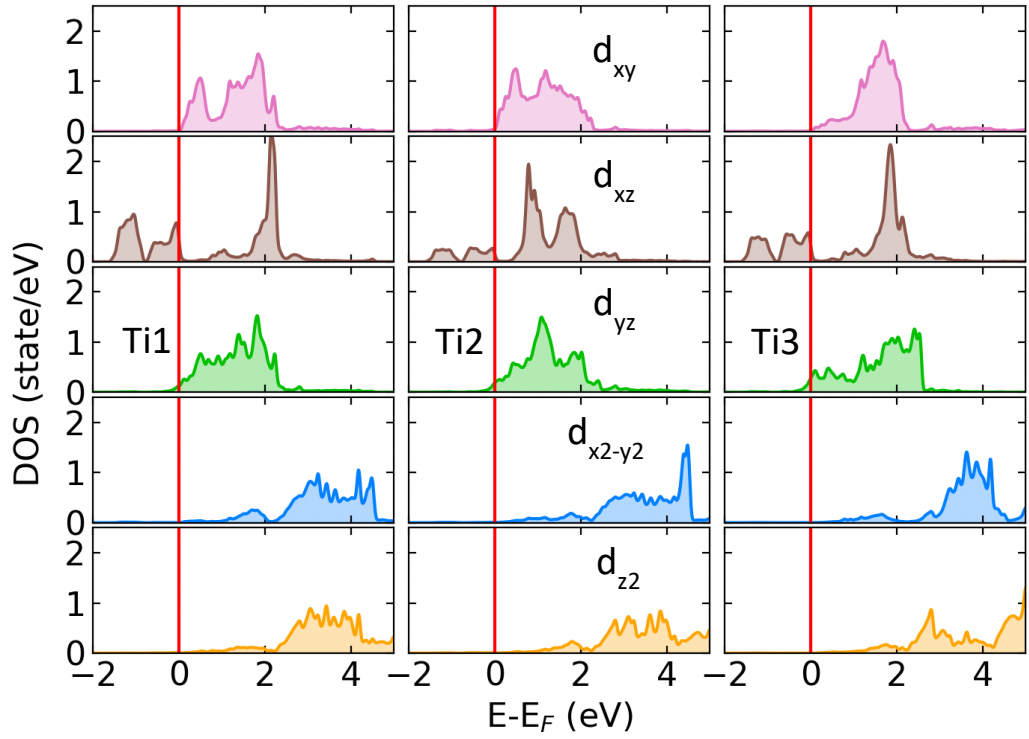

**Fig. S10.** Orbital-projected density of states for Ti$_1$, Ti$_2$ and Ti$_3$ atoms calculated in $\beta$-Ti$_3$O$_5$ phase at pressure of 15 GPa.

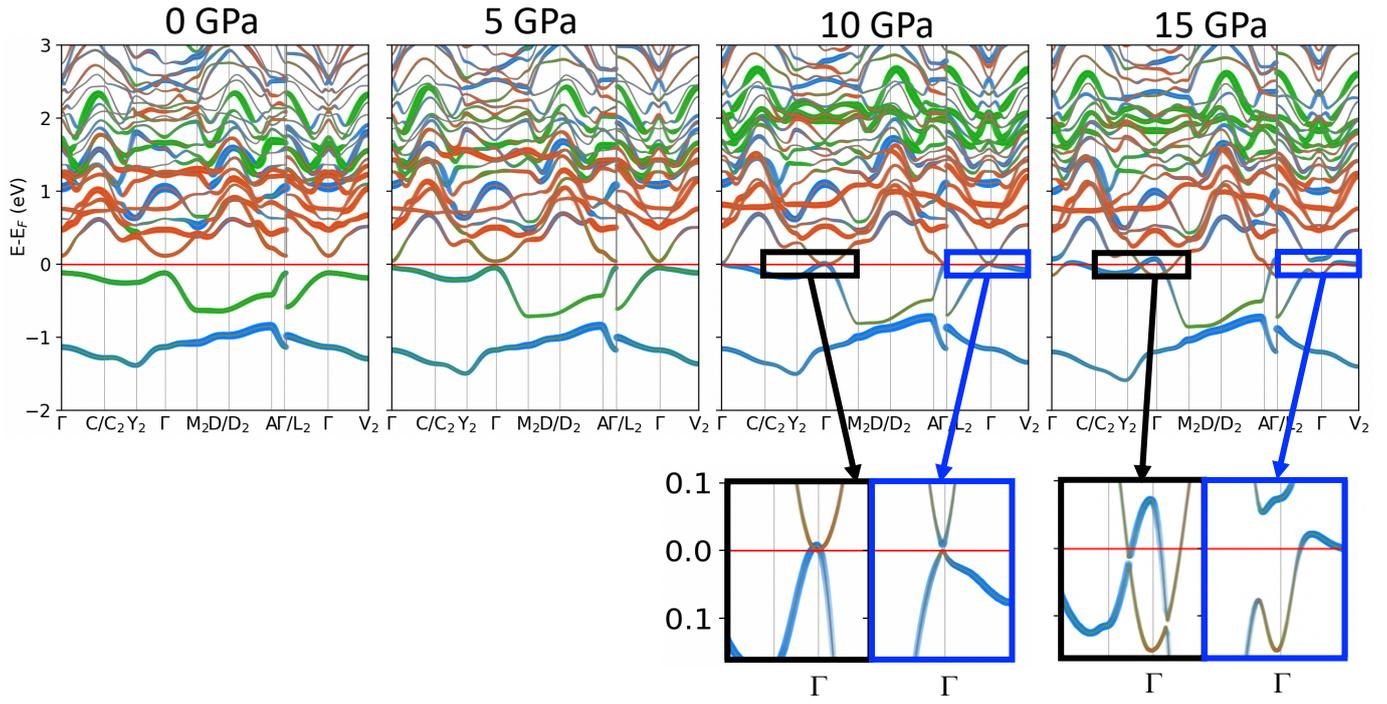

**Fig. S11.** Bands around the Fermi energy $E_F$ calculated in $\beta$-Ti$_3$O$_5$ as a function of external pressure. Insets are zoom-in at the $\Gamma$ point.